\documentclass[%
 reprint,
superscriptaddress,
 amsmath,amssymb,
pra,
]{revtex4-1}
\usepackage[utf8]{inputenc}
\usepackage{amsmath}
\usepackage{graphicx}
\usepackage{subcaption} 
\usepackage{dcolumn}
\usepackage{bm}
\usepackage{braket}
\usepackage[usenames,dvipsnames,svgnames,table]{xcolor}
\usepackage{hyperref}

\newtheorem{definition}{\indent Definition}
\newtheorem{Proposition}{\indent Proposition}

\newcommand{\textcal}[1]{$\mathcal{#1}$}
\newcommand{\pExp}{\mathcal{P}_{\not\in}}
\newcommand{\multinom}[2]{\begin{pmatrix}#1\\#2\end{pmatrix}}

\graphicspath{{fig/}}
\date{\today}

\begin{document}
\title{Many-box Locality}
\author{Yuqian Zhou}
\affiliation{State Key Laboratory of Networking and Switching Technology, Beijing University of Posts and Telecommunications, Beijing, 100876, China}
\affiliation{Centre for Quantum Technologies, National University of Singapore, Singapore}
\affiliation{State Key Laboratory of Information Security, Institute of Information Engineering, Chinese Academy of Sciences, Beijing, 100093, China}
\author{Yu Cai}
\affiliation{Centre for Quantum Technologies, National University of Singapore, Singapore}
\author{Jean-Daniel Bancal}
\affiliation{Quantum Optics Theory Group, University of Basel, Switzerland}
\author{Fei Gao}
\affiliation{State Key Laboratory of Networking and Switching Technology, Beijing University of Posts and Telecommunications, Beijing, 100876, China}
\author{Valerio Scarani}
\affiliation{Centre for Quantum Technologies, National University of Singapore, Singapore}
\affiliation{Department of Physics, National University of Singapore, Singapore}
\begin{abstract}
    There is an ongoing search for a physical or operational definition for quantum mechanics. Several informational principles have been proposed which are satisfied by a theory less restrictive than quantum mechanics. Here, we introduce the principle of ``many-box locality'', which is a refined version of the previously proposed ``macroscopic locality''. These principles are based on coarse-graining the statistics of several copies of a given box. The set of behaviors satisfying many-box locality for $N$ boxes is denoted $MBL_N$. We study these sets in the bipartite scenario with two binary measurements, in relation with the sets $\mathcal{Q}$ and $\mathcal{Q}_{1+AB}$ of quantum and ``almost quantum'' correlations. We find that the $MBL_N$ sets are in general not convex. For unbiased marginals, by working in the Fourier space we can prove analytically that $MBL_{N}\subsetneq\mathcal{Q}$ for any finite $N$, while $MBL_{\infty}=\mathcal{Q}$. Then, with suitably developed numerical tools, we find an example of a point that belongs to  $MBL_{16}$ but not to $\mathcal{Q}_{1+AB}$. Among the problems that remain open, is whether $\mathcal{Q}\subset MBL_{\infty}$.
\end{abstract}

\maketitle

\section{Introduction}

The definition of quantum physics is most frequently reduced to a description of its mathematical formalism: physical systems are described by vector spaces, and their properties by subspaces. The desire for a more physical, or operational, or even philosophical foundation for this definition is an ongoing task. Two programs have reported significant advances in the last decade.

The first program is a revival of the attempts left pending in the approach called ``quantum logic''. Quantum physics is put in the context of generalized probabilistic theories, then singled out through a small set of axioms. Breakthrough was achieved by realizing that one needs axioms about composite systems, i.e. one that captures some aspects of entanglement. This program has achieved the goal of reconstructing the Hilbert space structure (see~\cite{chiri} for a review).

Another program has been inspired by the work on Bell nonlocality. There, the basic mathematical object are the correlations among the outcomes of measurements on separated systems. One could try and find a physical principle that would allow exactly the set of correlations predicted by quantum theory. The pioneering attempt in this direction was that of Popescu and Rohrlich~\cite{PRbox}, who asked whether No-Signaling could be such a principle and found it defines a much larger set of correlations. This No-Signaling set became then the arena, in which the Quantum set had to be recovered by further constraints. The main principles proposed to date, inspired either by information theory or by physics, are Non-trivial Communication Complexity~\cite{NCC}, No Advantage for Nonlocal Computation~\cite{NANC}, Information Causality (IC)~\cite{IC}, Macroscopic Locality (ML)~\cite{ML} and Local Orthogonality~\cite{LO}. Each defines a set of correlations which touches the quantum set in non-trivial way. For most of them, we don't have a compact characterization but (pending a general proof for IC) we know that they are larger than a set that is strictly larger than the quantum set~\cite{almostQuantum}. In other words, the principle defining quantum physics in this program is still being sought \cite{hypersignal,almostnew}.

\begin{figure}
    \centering
    \begin{subfigure}[b]{0.4\textwidth}
        \centering
        \includegraphics[width=0.5\textwidth]{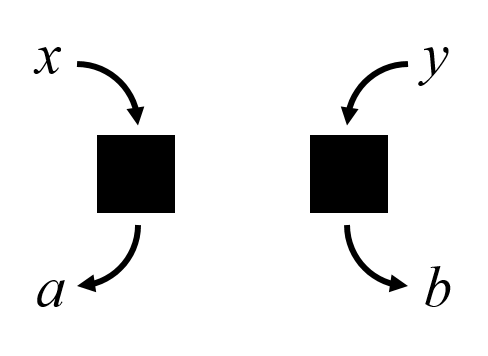}
        \caption{The usual Bell scenario. We called this a ``box".}
        \label{fig:oneBox}
    \end{subfigure}
    
    \begin{subfigure}[b]{0.4\textwidth}
        \centering
        \includegraphics[width=0.6\textwidth]{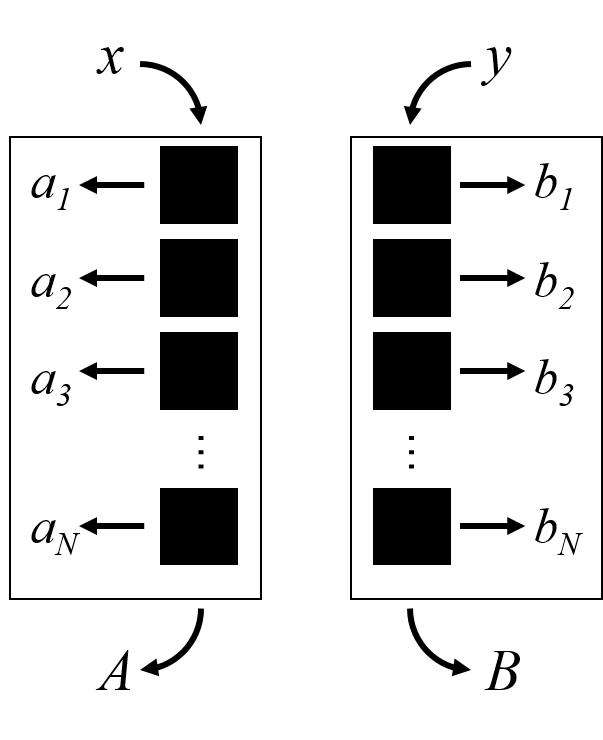}
        \caption{The many-box scenario. An ``$N$-box'' is formed by combining $N$ boxes together in a particular way: all individual boxes are identical and independent, the same setting is provided to each box, and the outcome of each box is cumulated to produce the outcome of the many-box, i.e. $A=\sum_i a_i$, $B=\sum_i b_i$.}
        \label{fig:nBox}
    \end{subfigure}
    \caption{Comparison between the usual Bell scenario and the many-box scenario.}
    \label{fig:setup}
\end{figure}

In this paper we explore the principle of Many-box locality (MBL) that is a refinement of ML. ML is probably the most physical of the principles listed above. The starting point is the fact that we don't see violation of Bell inequalities in the macroscopic world. The formalization considers that one can't observe the outcome of individual sources of correlations (\textit{boxes}), but only the coarse-graining of the outcomes of $N$ of them (see Fig.~\ref{fig:nBox}). Then, a second coarse-graining is considered: the outcomes of $N$ boxes are known with a precision $\sqrt{N}$. Under these coarse-graining assumptions, one can characterize exactly the set of correlations identified by ML in the limit $N\rightarrow \infty$: it coincides with the first step of the Navascu\'es-Pironio-Ac\'{\i}n hierarchy of semi-definite relaxations~\cite{NPA}. A similar work by Rohrlich~\cite{rohrlich2014} shows that in the infinite $N$ limit, some nonlocal boxes can even be activated to signal if weak measurements are possible for these boxes. By forbidding this possibility of activation of signalling, the Tsirelson's bound is recovered.  

The second coarse-graining is certainly very reasonable in practice, but one may question whether it fits in a fundamental principle: we may never be able to distinguish $10^{23}$ from $10^{23}+1$, but nature may not be defined by our limitations. MBL is defined by keeping the first coarse-graining of ML, but not the second one. Besides, we shall discuss the set of achievable correlations also for finite $N$, whereas ML was directly phrased for the infinite limit.

\section{Preliminary notions}

In this section, we first introduce the Bell scenario. Then three sets of correlations of interest will be introduced: the no-signalling set \textcal{NS}, the quantum set \textcal{Q}, the local set \textcal{L}, as well as Bell inequalities. For simplicity, we introduce these notions in bipartite scenarios, but these can be generalized to any general Bell scenarios including multipartite ones. 

\subsection{Bell scenario and behaviors}

Consider two parties Alice and Bob at distinct locations. Each has a measurement device, which shall be treated as a black box with the input $x\in\mathcal{X}=\{1,2,\cdots,M_{A}\}$ and $y\in\mathcal{Y}=\{1,2,\cdots,M_{B}\}$, and the output $a\in\mathcal{A}=\{1,2,\cdots,m_{A}\}$ and $b\in\mathcal{B}=\{1,2,\cdots,m_{B}\}$, for Alice and Bob respectively. We refer to this as the $(M_{A},M_{B},m_{A},m_{B})$ \textit{Bell scenario}.

In each run of the experiment, each party chooses a input at random and obtains an outcome. After repeating the experiment sufficiently many times, one can reconstruct the family of $M_{A}\times M_{B}$ probability distributions
\begin{equation*}
  \mathcal{P}_\mathcal{{X, Y}}=\{P(a, b|x, y),a\in\mathcal{A},b\in\mathcal{B}\}_{x\in\mathcal{X}, y\in\mathcal{Y}}
\end{equation*} with $P(a,b|x,y)\geq 0$ for all $a,b,x,y$ and $\sum_{a,b} P(a,b|x,y) = 1$ for all $x,y$. It has become customary to refer to the $\mathcal{P}_\mathcal{{X, Y}}$ as to a \textit{behavior}. For the purpose of this paper, we need to define three sets of behaviors: no-signaling ($\mathcal{NS}$), quantum ($\mathcal{Q}$) and local ($\mathcal{L}$). They are strictly included into one another according to $\mathcal{L}\subsetneq \mathcal{Q}\subsetneq \mathcal{NS}$. In this paper, we will be focusing on $(2,2,d,d)$ scenarios, though the notions in this section applies to any Bell scenarios. 

\subsection{The no-signaling set}

A behavior $\mathcal{P}$ is said to be no-signaling (NS) if it satisfies
\begin{equation}\label{defns}
\begin{array}{lcl}
  \sum_{b\in\mathcal{B}}P(a,b|x,y)\equiv P(a|x,y)&=& P(a|x)\\
  \sum_{a\in\mathcal{A}}P(a,b|x,y)\equiv P(b|x,y)&=& P(b|y),
  \end{array}
\end{equation}
for all $x,y$. The set of no-signalling behaviors $\mathcal{NS}$ is defined by a finite number of linear constraints, namely the positivity constraints $0 \leq P(a,b|x,y)\leq 1$, the normalization constraints $\sum_{a,b}P(a,b|x,y)=1$ and the no-signaling constraints \eqref{defns}. Because of this, it has a compact characterization: it is a \textit{polytope}, i.e. a convex set with finitely many extremal points.

Let us represent no-signaling behaviors in the $(2,2,d,d)$ Bell scenario in the Collins-Gisin form~\cite{collinsgisin}:
\begin{equation}
\mathcal{P}_\mathcal{{X,Y}}:=
\left(
 \begin{array}{c||c|c}
1 &  P(b|y=0) & P(b|y=1) \\ \hline\hline
P(a|x=0) & P(a,b|0,0) & P(a,b|0,1) \\ \hline
P(a|x=1) & P(a,b|1,0) & P(a,b|1,1) 
\end{array}
\right)\nonumber
\end{equation}
where $a,b\in\{0,1,\cdots,d-2\}$. Notice that a $(2,2,d,d)$ no-signaling behavior is fully specified by $4d(d-1)$ numbers, whereas a generic behavior in the same scenario would require $4(d^2-1)$ numbers.

\subsection{The quantum set}

For a given bipartite Bell scenario, the \textit{quantum set} $\mathcal{Q}$ of behaviors is the set of all the $\mathcal{P}_\mathcal{{X,Y}}$ that can be obtained with quantum theory. That is: $\mathcal{P}_\mathcal{{X,Y}}\in\mathcal{Q}$ if there exists a quantum
state $|\psi\rangle$, and two sets of measurement operators $\{E_{A}^{a|x}, a\in\mathcal{A}, x\in\mathcal{X}\}$ and $\{E_{B}^{b|y},b\in\mathcal{B}, y\in\mathcal{Y}\}$, such that $[E_{A}^{a|x}, E_{B}^{b|y}]=0$, $\sum_a E_{A}^{a|x}=\mathbb{I}$, $\sum_b E_{B}^{b|y}=\mathbb{I}$, and
\begin{equation}
 P(a, b|x, y) = \langle\psi|E_{A}^{a|x}E_{B}^{b|y}|\psi\rangle
\end{equation} for all $a,b,x,y$. Notice that, since no restriction is made on the dimension of the Hilbert space, the state can be taken to be pure and the measurement operators projective.

The quantum set is convex but not a polytope. So far, no compact characterization of the quantum set is known. In this paper, we shall use the hierarchy of semi-definite programs proposed by Navascu\'{e}s-Pironio-Ac\'{\i}n (NPA)~\cite{NPA} as an outer approximation of \textcal{Q}. Each level $j$ of the hierarchy defines a set of behaviors larger than the quantum set, such that $\mathcal{Q}_{j+1}\subseteq \mathcal{Q}_{j}$; the hierarchy converges to the quantum set, i.e. $\mathcal{Q}_{j\rightarrow\infty}=\mathcal{Q}$.

\subsection{The local set}

A local deterministic behavior is one in which the outputs $a$ and $b$ are determined by the inputs $x$ and $y$, respectively:
\begin{equation}
 P_{LD}(a,b|x,y)=\delta_{a=f_{i}(x)}\delta_{b=f_{j}(y)},
\end{equation} There are $m_A^{M_A}\,m_B^{M_B}$ local deterministic behaviors. As an example that will be used later, in the $(2,2,2,2)$ scenario, the behavior corresponding to deterministically outputting $a=1$ and $b=1$ for both inputs is represented by
\begin{equation}
 \mathcal{P}_{LD_1}=
\left(
 \begin{array}{c||c|c}
1 &  0 & 0 \\ \hline\hline
0 & 0 & 0 \\ \hline
0 & 0 & 0 
\end{array}
\right).
\end{equation}

A behavior $\mathcal{P}$ is said to be local if and only if can be described as a \textit{convex combination of local deterministic behaviors}:
\begin{eqnarray}
\mathcal{P}&=&\sum_i c_i \mathcal{P}_{LD_i},
\end{eqnarray}
where $ c_i \geq 0,\sum_i c_i=1$. Thus, the local set $\mathcal{L}$ is also a polytope, with the local deterministic behaviors as extremal points.

The local polytope shares some facets with $\mathcal{NS}$, defined by the positivity, normalization and no-signaling constraints. Its proper facets are \textit{Bell inequalities} $C$:
\begin{eqnarray}
  C\cdot \mathcal{P}\equiv\sum_{a,b,x,y} C(a,b,x,y)P(a,b|x,y)\,\leq V
\end{eqnarray}
where $V$ is called the local value. In other words, a behavior is local if and only none of the Bell inequalities is violated. To test whether a given point is local, it is not necessary to list down all the Bell inequalities explicitly for that scenario. Rather, one can solve the following linear program:
\begin{eqnarray}\label{linopt}
\text{maximise\ } &\quad &v\\
\mbox{}\text{subject to} &\quad & \mathcal{P}(v)=v\mathcal{P}+(1-v)\mathcal{P}_{mix},\nonumber\\
&\quad& \mathcal{P}(v)=\sum_i t_i \mathcal{P}_{LD_i}, \nonumber\\ 
&\quad& t_i \geq 0,\sum_i t_i=1,\nonumber
\end{eqnarray}
where $\mathcal{P}_{mix}(a,b|x,y)=1/d^2, \forall a,b,x,y$, is the maximally mixed distribution. We call the returned value $v_{max}$ the local parameter. The behavior $\mathcal{P}$ is local if $v_{max}\geq1$; non-local if $v_{max}<1$. As a by-product of this optimization, a certificate is provided in either case. If the test distribution is local, it will return us a valid decomposition in terms of local deterministic points. Otherwise, it will return a Bell inequality that is violated by the test distribution as a consequence of Farkas' lemma~\cite{FL}. 

\section{Many-box locality: definition and tools}

In this section, we first introduce the notion of many-box coarse graining and many-box locality (MBL). Second we introduce the tool of Fourier transform that would be useful for this study.

\subsection{Definition of many-box locality}

Here for simplicity, we will define many-box locality in the $(2,2,2,2)$ scenario, though it can be generalized to any Bell scenario, including multipartite ones. Consider $N$ identical $(2,2,2,2)$ boxes, each described by $\mathcal{P}(a,b|x,y)$. In any run, all the boxes take the same inputs $x$ and $y$, but each produces independently the output $a_i,b_i \in \left\{ 0,1 \right\}$, $i=1,\cdots,N$. Besides, we do not keep all the information on the output, but just the locally coarse-grained variables $A=\sum_i a_i$ and $B = \sum_i b_i$. By construction, $A\in\left\{0,1,\cdots, N\right\}$ and $B\in\left\{0,1,\cdots, N\right\}$. In other words, the coarse-grained variables define a $(2,2,N+1,N+1)$ scenario.

The behavior of these variables will be denoted by $\mathcal{P}^{*N}$, called the \textit{$N$-box coarse-graining of $\mathcal{P}$}. It is not difficult to write down the explicit form of $\mathcal{P}^{*N}$ given $P(a,b|x,y)$: 

\begin{eqnarray}
 P^{*N}(A,B|x,y)&=&\sum_{k=0}^N \multinom{N}{A-k,B-k,k,N-A-B+k} \nonumber\\
 & &\cdot P(00|x,y)^{N-A-B+k}\cdot P(01|x,y)^{B-k} \nonumber\\
 & &\cdot P(10|x,y)^{A-k}\cdot P(11|x,y)^k \end{eqnarray}
where $\multinom{N}{n_1,n_2,n_3,n_4}=\dfrac{N!}{n_1!n_2!n_3!n_4!}$. In particular, the marginals are given by
\begin{eqnarray}
 P^{*N}(A|x)&=&\multinom{N}{A}P(a=1|x)^A\cdot P(a=0|x)^{N-A},\nonumber\\
 P^{*N}(B|y)&=&\multinom{N}{B}P(b=1|x)^B\cdot P(b=0|x)^{N-B},\nonumber
\end{eqnarray}

We introduce the notion of \emph{many-box local} (MBL) sets:
\begin{definition}
A behaviour $\mathcal{P}$ is said to be $N$-box local $(\mathcal{P}\in MBL_N)$ if $\mathcal{P}^{*N}$ is local in the $(2,2,N+1,N+1)$ scenario. Notice that the sets $MBL_N$ are defined in the original scenario $(2,2,2,2)$.
\end{definition}

By construction, the coarse-graining procedure obeys the composition rule
\begin{eqnarray}
 \mathcal{P}^{*(N_1+N_2)}=\mathcal{P}^{*N_1}*\mathcal{P}^{*N_2}\,.
\end{eqnarray} From it, one can straightforwardly prove the following inclusion relations:

\begin{Proposition} If $\mathcal{P}\in \bigcap_{j} MBL_{N_j}$, then $\mathcal{P}\in MBL_{N(\vec{q})}$ with $N(\vec{q})=\sum_j q_jN_j$, for all $\vec{q}=(q_1,q_2,...)$ with $q_j\in\mathbb{N}$.  
\end{Proposition}

Indeed, local decompositions for both $\mathcal{P}^{*N_1}$ and $\mathcal{P}^{*N_2}$ directly define a local decomposition for $\mathcal{P}^{*(N_1+N_2)}$. As a corollary: if a probability distribution is both $2$-box and $3$-box local, then it is $N$-box local for any $N\geq 2$, since any $N$ can be decomposed as a sum of multiples of two and three.

Beyond this Proposition, we have not been able to find general properties for the $MBL_N$ sets. Properties that might have been conjectured, e.g. convexity or the inclusion $MBL_{N}\subseteq MBL_{N+1}$, do not hold, as forthcoming counterexamples will demonstrate.

\subsection{Fourier transform}
\label{sec:fourier}
The coarse-grained behaviors $\mathcal{P}^{*N}$ are obtained by \textit{convolution} of $N$ copies of $\mathcal{P}$. This observation suggests studying the problem in its Fourier transformed version. To the best of our knowledge, this is a new method in non-locality studies.

First, we define the Fourier transform of a behavior:
\begin{definition}
The Fourier transform of order $r$, where $r\geq d$, of a behaviour $\mathcal{P}$, denoted $\mathcal{F}_r[
\mathcal{P}]$, is defined as
\begin{eqnarray}
 \mathcal{F}_r[\mathcal{P}](k,l|x,y)&=&\tilde{P}_r(k,l|x,y)\nonumber\\
 &:=&\sum_{a, b=0}^{r-1} e^{\frac{2\pi i}{r}(ak+bl)}P(a,b|x,y),\nonumber
\end{eqnarray}
for any $x,y$, where $k,l \in\{0,1, \cdots, r-1\}$.
\end{definition}

Similarly, we define the inverse Fourier transform of a Bell expression $C$:
\begin{definition}
The inverse Fourier transform of order $r$, where $r\geq d$, of a Bell expression $C$, denoted $\mathcal{F}^{-1}_r[C]$, is defined as
\begin{eqnarray}
\mathcal{F}_r^{-1}[C](k,l,x,y)&=& \tilde{C}_r(k,l,x,y)\nonumber\\
&:=&\frac{1}{r^2}\sum_{a, b=0}^{r-1} e^{-\frac{2\pi i}{r}(ak+bl)}C(a,b,x,y),\nonumber
\end{eqnarray}
for any $x,y\in \{0,1\}$, where $k,l \in\{0,1, \cdots, r-1\}$.
\end{definition}

Note that the Fourier transform is only well-behave when $r \geq d$. So this is always assumed for the rest of the paper and the subscript $r$ is dropped for simplicity. 

Due do the linearity of Fourier transform, Bell violation is preserved, that is:
\begin{eqnarray}\label{fourier trans}
 \mathcal{F}_r^{-1}[C]\cdot\mathcal{F}_r[\mathcal{P}]&=&
 \sum_{k,l,x,y}\tilde{C}_r(k,l,x,y)\tilde{P}_r(k,l|x,y)\nonumber\\
 &=&\sum_{a,b,x,y}C(a,b,x,y)P(a,b|x,y)\nonumber\\
 &=& C\cdot \mathcal{P}.
\end{eqnarray}

Another property of the Fourier transform follows from the convolution theorem, applied repeatedly:
\begin{eqnarray}
\mathcal{F}_r[\mathcal{P}^{*N}]=(\mathcal{F}_r [\mathcal{P}])^N\,.
\end{eqnarray}
The $N$-box coarse-graining manifests as raising to the $N$-th power the Fourier transformed probability. 

We can now analyze many-box locality in the Fourier transformed space. $\mathcal{P}^{*N}$ is local if and only if it satisfies all the Bell inequalities in the $(2,2,N+1,N+1)$ scenario, $C_i \cdot \mathcal{P}^{*N} \leq V_i$. By linearity and the convolution theorem, for the Fourier transformed probability, we have
\begin{align}
    \label{eqn:fourierViolation}
    \tilde{C}_i \cdot \widetilde{\mathcal{P}^{*N}} = \tilde{C}_i \cdot \widetilde{\mathcal{P}}^{N} \leq V_i,
\end{align}
for all $i$. Note that each Bell inequality $(C_i,V_i)$ defines an $N$-th degree polynomial inequality for single box behavior $\mathcal{P}$. Thus on the boundary of $MBL_N$, one or more such polynomial inequalities is saturated. We summarize this in the following:

\begin{Proposition}
Let $\mathcal{R}_i^N:= \big\{\mathcal{P}\,|\,\tilde{C}_{i}\cdot\tilde{P}^N\leq V_i\big\}$. Then $MBL_N=\bigcap_{i\in I}\mathcal{R}_i^N$, where $\{(C_i,V_i)\}_ {i\in I}$ is the set of all the Bell inequalities of the $(2,2,N+1,N+1)$ scenario. Behaviors on the boundary of $MBL_N$ must saturate one or several Bell inequalities: 
\begin{align}
 \mathcal{P} \in bd(MBL_N) \Rightarrow \exists i\in I, \text{ such that } \tilde{C}_{i}\cdot\tilde{\mathcal{P}}^N=V_i.\nonumber
\end{align}
\end{Proposition}
This provides a systematic way to compute the boundary of $MBL_N$. However this is not tractable for large $N$ due to the (at least) exponentially increasing number of Bell inequalities in the $(2,2,N+1,N+1)$ scenario. The task is now to find the relevant set of Bell inequalities that defines the border of the $MBL_N$ sets. As we will see in the following sections, in some cases we could guess the relevant set of Bell inequalities, hence characterizing the $MBL_N$ set; otherwise, any particular set of Bell inequalities will provide an upper bound to the $MBL_N$ set.

\section{Results on $MBL_N$ in the $(2,2,2,2)$ scenario}

Now we address explicitly the characterisation of $MBL_N$ in the $(2,2,2,2)$ Bell scenario. We start by plotting some of the $MBL_N$ in some slices, by solving the linear optimisation \eqref{linopt}. These plots show that the shape of the sets and their inclusion is not trivial. Then we present one case in which the tool of Fourier transforms can be used to find the boundary of $MBL_\infty$ in a special slice. Finally we discuss the relation between MBL and the sets $\mathcal{Q}_n$ of the NPA hierarchy, including $\mathcal{Q}_1$ that defines ML.

\subsection{Numerical plots}

To have a better understanding of these $MBL_N$ sets, in this section, we start with numerical plots of $MBL_N$ sets. We first show $MBL_2$ and $MBL_3$ in a three dimensional slice of the \textcal{NS} polytope. Then, we show $MBL_N$ for $N$ up to $10$ in a two-dimensional slices. 

The local polytope in the $(2,2,2,2)$ scenario is characterized by its 16 extremal points, $\mathcal{P}_{LD_{i}}, i=1,2,\cdots,16$. Its nontrivial facets consists of 8 CHSH~\cite{CHSH} inequalities, for example
\begin{eqnarray}
 C_1&\equiv& E_{00}+E_{01}+E_{10}-E_{11}\leq 2,\nonumber\\
 C_2&\equiv& -E_{00}+E_{01}+E_{10}+E_{11}\leq 2,
\end{eqnarray}
where
\begin{equation}
 E_{xy}=P(a=b|x,y)-P(a\neq b|x,y).\nonumber
\end{equation}
No-signalling probability distribution that violates each CHSH inequality to its algebraic maximum is known as a Popescu-Rolich-box (PR-box). For the two CHSH mentioned above, we have the corresponding PR-boxes:
\begin{equation}
 \mathcal{P}_{PR_1}=
\left(
 \begin{array}{c||c|c}
1 &  1/2 & 1/2 \\ \hline\hline
1/2 & 1/2 & 1/2 \\ \hline
1/2 & 1/2 & 0 
\end{array}
\right),
\end{equation}
and
\begin{equation}
 \mathcal{P}_{PR_2}=
\left(
 \begin{array}{c||c|c}
1 &  1/2 & 1/2 \\ \hline\hline
1/2 & 0 & 1/2 \\ \hline
1/2 & 1/2 & 1/2 
\end{array}
\right).
\end{equation}


We are going to parametrize the slice under consideration with $\mathcal{P}_{PR_1}, \mathcal{P}_{PR_2}, \mathcal{P}_{LD_1}$, and $\mathcal{P}_{mix}$:
\begin{eqnarray}
 & &\mathcal{P}(\alpha,\beta,\gamma)\\
 &=&\alpha \mathcal{P}_{PR_1}+\beta \mathcal{P}_{PR_2}+\gamma \mathcal{P}_{LD_1}+(1-\alpha-\beta-\gamma)\mathcal{P}_{mix}\nonumber\\
 &=& \left(
 \begin{array}{c||c|c}
1 &  (1-\gamma)/2 & (1-\gamma)/2 \\ \hline\hline
(1-\gamma)/2 & (1+\alpha-\beta-\gamma)/4 & (1+\alpha+\beta-\gamma)/4 \\ \hline
(1-\gamma)/2 & (1+\alpha+\beta-\gamma)/4 & (1-\alpha+\beta-\gamma)/4 
\end{array}
\right).\nonumber
\end{eqnarray}

Thanks to the symmetry of PR-boxes, we only need to discuss the case where $\alpha, \beta \geq 0$. We first consider the region $\gamma \geq 0$. The positivity constraint is given by $\alpha + \beta + \gamma \leq 1$. The local polytope is constrained by two CHSH inequalities, $2\alpha+\gamma\leq 1$ and $2\beta+\gamma\leq1$. The quantum set is well characterized for the planes, $\alpha=0$, $\beta=0$ and $\gamma = 0$, by $\sqrt{2}\beta+\gamma=1$, $\sqrt{2}\alpha+\gamma=1$, and $\alpha^2+\beta^2=1/2$ respectively (the first two by~\cite{KT} and the last is the well known Tsirelson-Landau-Masanes arcsine inequality~\cite{Tsirelson,Landau,quancor}). We conjecture that in the region $\alpha,\beta,\gamma >0$, the quantum set is given by 
\begin{align}
 \alpha^2+\beta^2 \leq (1-\gamma)^2/2.
\end{align}


\begin{figure}
    \includegraphics[trim={0cm 4.2cm 0cm 4cm},clip,width=0.48\textwidth]{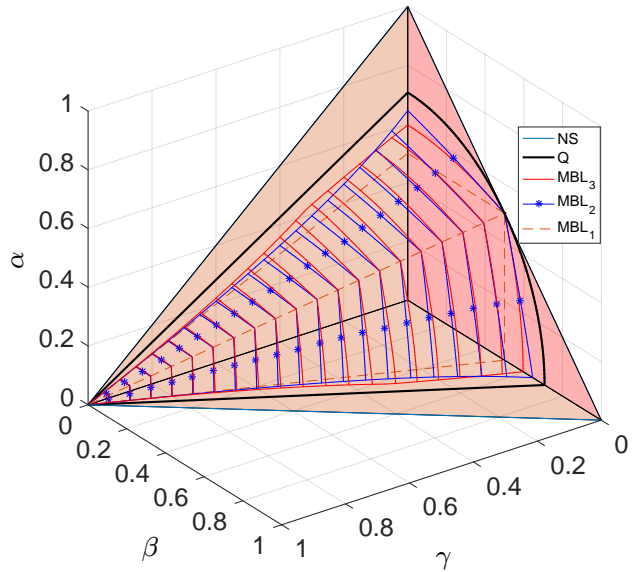}
    \caption{(Color online) The $MBL_1$, $MBL_2$ and $MBL_3$ in the No-signalling set.
    The region inside the dashed brown lines is the set $MBL_1$, which is the local set. Here, the thick black, the blue with star markers and the red lines are the boundary of $Q$, $MBL_2$ and $MBL_3$ respectively.}
    \label{fig:3d}
\end{figure}

In figure~\ref{fig:3d}, $MBL_1$, $MBL_2$ and $MBL_3$ are plotted for the region $\alpha,\beta,\gamma \geq 0$, as well as \textcal{N}, \textcal{Q} and \textcal{L}. By definition, $MBL_1$ is simply the local set \textcal{L}. $MBL_2$ and $MBL_3$ are obtained by solving the optimization~\eqref{linopt}. We would like to highlight two qualitative features about these MBL sets. First, $MBL_N$ sets are not convex as we can see from the figure. Mathematically, it says that to show $\mu \mathcal{P}_1 + (1-\mu)\mathcal{P}_2 \in MBL_N$, knowing $\mathcal{P}_1 \in MBL_N$ and $\mathcal{P}_2 \in MBL_N$ is not sufficient. Second, in general $MBL_N$ is not contained in $MBL_{N+1}$, as we can see in the figure, $MBL_2$ is not contained in $MBL_3$. For example, $\mathcal{P}(\alpha) = \mathcal{P}(\alpha,0,0)$ is in $MBL_2$ but not $MBL_3$ for $ \sqrt{7}-2 < \alpha \leq (1+\sqrt{2})^{1/3}-(1+\sqrt{2})^{-1/3}$.

\begin{figure}
 \includegraphics[trim={0cm 6cm 0cm 4.2cm},clip,width=0.48\textwidth]{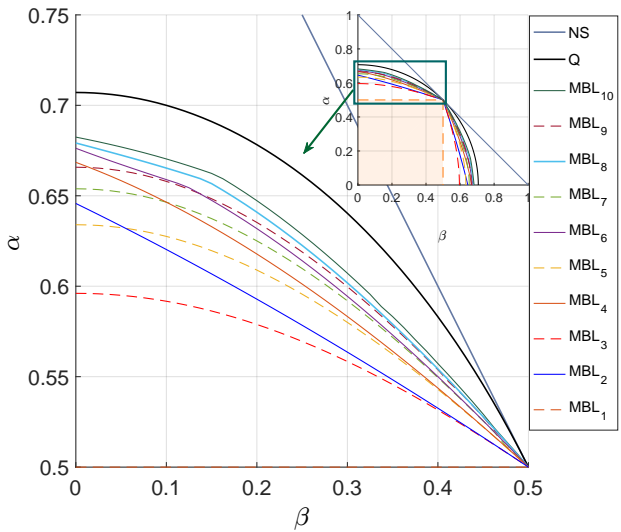}
  
    \caption{(Color online) The boundary of the $MBL_N$ sets  in the plane $\gamma=0$ for $N$ from 2 to 10. Here, solid lines from top to bottom are $NS$, $Q$, $MBL_{10}$, $MBL_8$, $MBL_6$, $MBL_4$, $MBL_2$, respectively, and dashed lines from top to bottom are $MBL_9$, $MBL_7$, $MBL_5$, $MBL_3$, $MBL_1$, respectively. Note that the shaded is the set $MBL_1$, i.e. the local set. In particular, we magnify the region inside the green rectangle box to show more details.} 
    \label{fig:gamma=0}
\end{figure}

\begin{figure}
    \includegraphics[trim={0cm 10cm 0cm 6cm},clip,width=0.48\textwidth]{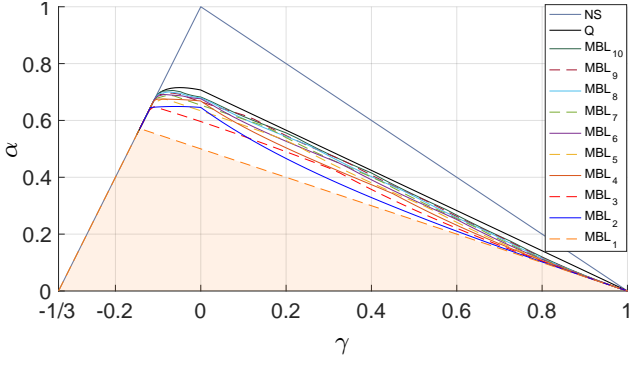}
    
    \caption{(Color online) The boundary of the $MBL_N$ sets  in the plane $\beta=0$ for $N$ from 2 to 10. Here, solid lines from top to bottom are $NS$, $Q$, $MBL_{10}$, $MBL_8$, $MBL_6$, $MBL_4$, $MBL_2$, respectively, and dashed lines from top to bottom are $MBL_9$, $MBL_7$, $MBL_5$, $MBL_3$, $MBL_1$, respectively.  Note that the shaded is the set $MBL_1$, i.e., the local set.} 
    \label{fig:beta=0}
\end{figure}

In figure~\ref{fig:gamma=0} and~\ref{fig:beta=0}, the $MBL_N$ sets are plotted in two 2d slices corresponding to $\gamma = 0$ and $\beta=0$. 
Note that for $\gamma < 0$, the positivity constraint becomes $\alpha + \beta -3\gamma \leq 1$ and the quantum boundary is well approximated by the second level of the NPA hierarchy, $\mathcal{Q}_2$. From the figures, we can see that for up to $N=10$, the $MBl_N$ are contained in the quantum set \textcal{Q}. The general trend of the sets are increasing with $N$. Though the inclusion relation $MBL_N \subseteq MBL_{N+1}$ is not true, $MBL_N \subseteq MBL_{N+2}$ is conjectured to hold for both even and odd $N$.

In the next section, we present some analytic results, especially for the case for $N \rightarrow \infty$. 

\subsection{$MBL_{\infty}=\mathcal{Q}$ for unbiased marginals}
\label{sec:analytical}

In figure~\ref{fig:gamma=0} in the previous section, the numerical plots for the boundary of the $MBL_N$ sets for $N$ odd are reasonably smooth. A closer inspection at the dual of the linear program~\eqref{linopt} reveals that the relevant inequality has a special form. For odd $N$ where the number of outcome $d=N+1$ is even, the relevant inequality that defines the boundary of $MBL_N$ is the so-called \emph{half-half} lifting~\cite{lifting} of CHSH. It is equivalent to first coarse graining the $d$ outcomes in to a binary outcome, such that Alice (Bob) assigns $+$ if $A\geq \frac{d}{2}$ ($B\geq \frac{d}{2}$), $-$ otherwise, then applying the CHSH inequality. Using the array notation for inequalities, expressed in terms of full probabilities
, we can also write:
        
\begin{eqnarray}
C_{half}&=&
\left(
 \begin{array}{r |r}
-\mathcal{J} &  \mathcal{J}\\ \hline
\mathcal{J} &  \mathcal{J}
\end{array}
\right),\,\nonumber
\mathcal{J}=
\left(
 \begin{array}{r r}
J & -J\\ 
-J &  J 
\end{array}
\right),
\end{eqnarray}
and $J$ is a matrix of ones of size $d/2\times d/2.$ 


Now let us consider the class of behaviors with fully random marginals, that is $P(a|x)=P(b|y)=\frac12$ for all $a,b,x,y$. Such a behavior can be parameterize by four parameters:

\begin{equation}
  \mathcal{P}_{half}= \left( \begin{array}{r |r}
P(\mu_1) & P(\mu_2)\\ \hline
P(\mu_3) & P(\mu_4)\nonumber
\end{array}
\right). 
\end{equation}
where
\begin{equation}
  P(\mu)= \frac{1}{4}\left( \begin{array}{r r}
1+\mu & 1-\mu\\ 
1-\mu & 1+\mu\nonumber
\end{array}
\right). 
\end{equation}

In the following, we aim to show that, in this slice of fully random marginals, the boundary defined by the half-half lifting, in the limit where $N$ tends to infinity, coincides with that of $\mathcal{Q}_1$. The Fourier transform introduced in section~\ref{sec:fourier} will be used in the proofs below. 

We start by computing the inner product of \textcal{J} and $P(\mu)$:
\begin{align}
    g(\mu) = \mathcal{J}\cdot P(\mu)^{*N}, \nonumber
\end{align}
which in the Fourier transformed space, following Eq.~\eqref{eqn:fourierViolation}, becomes:
\begin{align}
\label{eqn:mu}
    g(\mu) = \mathcal{F}^{-1}[\mathcal{J}] \cdot (\mathcal{F}[P(\mu)])^N.
\end{align}

Let us first work out the inverse Fourier transform of the Bell coefficient. Define the $d$-th root of unity as $\omega=e^{\frac{2\pi i}{d}}$, $R_+ = \{0,1,\cdots \tfrac d2-1\}$, and $R_- = \{\tfrac d2,\tfrac d2+1, \cdots, d-1\}$, we have:
\begin{align}
\label{eqn:fourierHalf}
 &\mathcal{F}_d^{-1}[\mathcal{J}](k,l)\nonumber\\
 &= \frac{1}{d^2}\Big(\sum_{(a,b)\in \substack{R_+ \times R_+ \\ R_-\times R_-}} \omega^{-(a k+bl)} 
 - \sum_{(a,b)\in \substack{R_+ \times R_- \\ R_-\times R_+}} \omega^{-(a k+bl)}\Big)\nonumber\\
 &=\frac{(1-\omega^{-\frac{dk}{2}})^2(1-\omega^{-\frac{dl}{2}})^2}{d^2(1-\omega^{-k})(1-\omega^{-l})}.
\end{align}

Then, we focus on the $\mathcal{F}_d^N[P(\mu)]$ term. Since $P(a,b|x,y)=0$ for $a,b > 1$, only four terms survives, 
\begin{align}
\mathcal{F}_d[P(\mu)](k,l)&=&\frac14\big[(1+\omega^k)(1+\omega^l)+(1-\omega^k)(1-\omega^l)\mu].
\end{align}


Now we are ready to expand Eq.~\eqref{eqn:mu} as a polynomial in $\mu$. After some calculations, we have
\begin{equation}
g(\mu)=\sum_{m=odd}^{N}C(d,m)\mu^m, \nonumber
\end{equation}
where we write $C(d,m)$ in terms of Gamma functions $\Gamma$:
\begin{equation}
C(d,m):=\frac{16\Gamma(d)}{4^d \Gamma(m+1) \Gamma(d-m)}\Bigg(\frac{\Gamma(d-m) \Gamma(m)}{\Gamma(\frac{d}{2})\Gamma(\frac{d-m+1}{2})\Gamma(\frac{m+1}{2})}\Bigg)^2.\nonumber
\end{equation}
Only coefficient for odd $m$ survives, $C(d,m)=0$ for even $m$ because that the denominator diverges to infinity, see Appendix~\ref{appa} for more detail. 

Furthermore, evaluating the limit when $N\rightarrow\infty$, one obtains 
\begin{equation}
\label{eq:doubleFactorial}
C(\infty,m)=\frac{2}{\pi}\frac{1}{m}\frac{(m-2)!!}{(m-1)!!},
\end{equation}
where $(\cdot)!!$ is the double factorial (see Appendix A).
Note that this resembles the Taylor expansion of $\arcsin(\mu)$, which reads:
\begin{equation}
\arcsin(\mu)=\sum_{m=odd}^\infty \frac{1}{m}\frac{(m-2)!!}{(m-1)!!}\mu^m.
\end{equation}

In conclusion, in the limit of $N$ going to infinity, testing the Bell inequality $C_{half} \cdot \mathcal{P}^{*N} \leq 2$ reduces to testing the following Bell inequality:
\begin{equation}
\label{eqn:arcsin}
\frac{2}{\pi}(-\arcsin(E_{00}) + \arcsin(E_{01}) + \arcsin(E_{10}) + \arcsin(E_{11})) \leq 2
\end{equation}
on the single box behavior. Incidentally, Eq.~\eqref{eqn:arcsin} also characterizes the $\mathcal{Q}_1$ set of behaviors. 




As a by-product, one can derive the analytic solution of the boundary of $MBL_N$, for odd $N$, in the plane $\gamma=0$. For example, the boundaries of $MBL_N$, $N=1,3,5$ are shown in the following Table.
\begin{table}[h!]
    \centering
    \begin{tabular}{c|c}
        $N$ & expression \\\hline\hline
        1 & $4\beta=2$\\
        3 & $\beta(3+3\alpha^2+\beta^2)=2$\\
        5 & $\beta(45+45\alpha^4+10\beta^2+9\beta^4+30\alpha^2(1+3\beta^2))/16=2$
    \end{tabular}
\end{table}
Note that the boundary in the case of $N=1$ is simply the CHSH-Bell inequality.

\subsection{A sequence of quantum behaviors outside $MBL_N$}

In this section, we present an analytic sequence of quantum behaviors that is $N$-nonlocal for each $N$. This family makes use of a different binning of the $N+1$ outcomes than the half-half binning in the previous section: instead,
we consider that each party outputs the parity of its outcome, e.g. $(-1)^A$ for Alice. We note that this binning requires a full resolution of the outcomes $A$ and $B$, hence it is not possible in the context of ML, where the resolution is limited~\cite{ML}.

Consider the probability distribution obtained from measuring the singlet state
\begin{align*}
    \ket{\psi} = \frac{1}{\sqrt{2}} (\ket{01} - \ket{10}),
\end{align*}
in the following basis
\begin{align*}
    A_0:& \;\sigma_Z, &&A_1: \cos(2\theta)\sigma_Z+\sin(2\theta) \sigma_X, \\
    B_0:& \cos(\theta)\sigma_Z + \sin(\theta) \sigma_X, &&B_1: \cos(\theta)\sigma_Z - \sin(\theta)\sigma_X.
\end{align*}
With these measurements, we can compute the four correlators for a single box:
\begin{align*}
    E_{00} = E_{01} = E_{10} = \cos(\theta), \; E_{11} = \cos(3\theta);
\end{align*}
Incidentally, this behavior always sits on the boundary of \textcal{Q}, since it satisfies the conditions in Ref.~\cite{quancor}.

One can test the nonlocality of its $N$-box coarse-graining by the so-called \emph{parity binning}~\cite{Poh17} of the CHSH inequality, denoted as $C'$, where
\begin{eqnarray}
C'&=&
\left(
 \begin{array}{r |r}
\mathcal{J'} &  \mathcal{J'}\\ \hline
\mathcal{J'} &  -\mathcal{J'}
\end{array}
\right),\,\nonumber
\mathcal{J'}_{i,j}=(-1)^{i+j}.
\end{eqnarray}

Following parity binning, and because the $N$ box are identical and independent, we can see that the correlation
\begin{align}
    \langle E^{(N)}_{x,y} \rangle &= \langle (-1) ^ {A_x+B_y}\rangle = \langle (-1)^ {\sum_i a_{x,i}+b_{y_i}}\rangle \nonumber \\
    & = \langle (-1)^{a_{x,i} + b_{y,i}}\rangle^N = E_{x,y}^N,
\end{align}
that is the correlation for the $N$-box coarse-graining is simply the $N$th power of that of a single box. Hence,

\begin{align}
\label{eqn:parity}
   S_N = C'\cdot \mathcal{P}^{*N}& =E_{00}^N + E_{01}^N + E_{10}^N - E_{11}^N \nonumber\\
   & =3\cos^N(\theta_N) - \cos^N(3\theta_N).
\end{align}


Let $\theta_N=\frac{\theta_0}{\sqrt{N}}$, this quantity can be larger than 2 for any $N$ with suitable choice of $\theta_0$. 
Moreover, in the limit of $N$ going to infinity
\begin{align*}
    S_N = 3 e^{-\theta^2_0/2} - e^{-(3\theta_0)^2/2},
\end{align*}
whose maximum $S_N^{\max} = 3^{7/8}-3^{-9/8}\approx 2.3245$ is achieved with $\theta_0 = \frac{\sqrt{\ln 3}}{2}$. 

It is worth emphasizing that the quantum behavior considered here is different for each value of $N$ (i.e. the angle $\theta_N$ depends on $N$). In particular, setting $N=\infty$ produces a local box which belongs to $MBL_N\ \forall N$. Therefore, this family of behaviors does not constitute an example of quantum behavior that is outside $MBL_\infty$.




\subsection{A non-quantum $MBL_{16}$ behavior}

In this section, we present a behavior that is not quantum but nevertheless belongs to the $MBL_{16}$ set: $\pExp\notin\mathcal{Q}$ and $\pExp \in MBL_{16}$. This point lies in the slice where $\beta = 0$, and is defined by
 \begin{align*}
     \alpha &=0.287569286421973 \\
     \gamma &=0.505748781260095.
 \end{align*}
 
First be reminded that the boundary of the quantum set is known analytically on this slice for  $\alpha,\gamma \geq 0$. It is given by the straight line $\gamma = \tfrac{1}{\sqrt{2}}(1-\alpha)$. One can thus easily check that $\pExp$ is not in the quantum set. A direct computation also demonstrate that this point does not belong to the set of almost quantum correlations: $\pExp\notin\mathcal{Q}_{1+AB}$.

In order to test whether this point belongs to the $MBL_{16}$ set, we solved the linear program \eqref{linopt} with $\pExp^{*16}$. This linear program involves 83521 extremal points in an 1088-dimensional space and is tractable with standard solvers. However, the local parameter $v$ turns out to be 1 up to machine precision. We thus developped a high precision linear programming solver to determine whether $v$ is smaller or larger than 1. This solver is available online in the latest version of YALMIP~\cite{refiner,yalmip}. Using this tool, we solve the linear program~\eqref{linopt} with 50 digits of precision and find that the local parameter is
\begin{align*}
    v = 1+7.4\times 10^{-19},
\end{align*}
with $t_i\geq 0$ (strictly), $\sum t_i = 1+9.8 \times 10^{-55}$ and $|\sum_i t_i \mathcal{P}_{LD_i}- (v\pExp^{*16}+(1-v)\mathcal{P}_{mix})|_1 \leq 8.8 \times 10^{-36}$. We consider this a convincing numerical proof of a valid decomposition, since $v$ is significantly larger than $1$ compare to the precision with which the constraints are satisfied. The full high precision primal solution to this linear program is provided in the ancillary files of the arxiv submission~\cite{ancillary}.

In conclusion, we have shown an explicit example of a distribution that lies in $MBL_{16}$ but outside \textcal{Q} and $\mathcal{Q}_{1+AB}$. Since $MBL_{16} \subseteq MBL_{\infty}$, this also shows that $\mathcal{Q} \neq MBL_{\infty}$ and $\mathcal{Q}_{1+AB} \neq MBL_{\infty}$.

\section{Discussion and outlook}

In this paper we introduced the set of many-box local behaviors $MBL_N$. Since the MBL principle is less coarse-grained than ML, this set is a priori stricter than the set of ML, $MBL_\infty\subseteq\mathcal{Q}_1$. We have shown in section \ref{sec:analytical} that $MBL_\infty=\mathcal{Q}$ in one slice in which $\mathcal{Q}_1=\mathcal{Q}$ also holds. Also, we found that $MBL_{16}$ contains behaviors which belong neither to $\mathcal{Q}$, nor to the larger set $\mathcal{Q}_{1+AB}$. This shows that $MBL_\infty\nsubseteq\mathcal{Q}_{1+AB}$. In particular, the principle of many-box locality does not identify the quantum set, either for finite or for infinite $N$.

There are two main open questions. On the one hand, we do not know if MBL is satisfied by all quantum behaviors, i.e. if $\mathcal{Q}\subsetneq MBL_\infty$. The opposite would mean that there exist quantum boxes that remain nonlocal even in the limit of infinitely many copies, in a way that is washed out by the second coarse-graining of ML. On the other hand, we do not know whether in some slices $MBL_\infty$ comes closer to the quantum set than $\mathcal{Q}_1$ or than $\mathcal{Q}_{1+AB}$. If $MBL_\infty$ were equal to $\mathcal{Q}_1$, the second coarse graining of ML does not play any role. If $MBL_\infty\supsetneq \mathcal{Q}_{1+AB}$, many-box locality is a strictly weaker principle than ``almost-quantum''.

Other technical features of the $MBL_N$ sets, suggested by numerical studies, remain conjectural. First, all the relevant inequality that detects the nonlocality of $N$-box distributions seem to be liftings of the CHSH inequality; we have not found an example where no liftings of CHSH is violated while a genuine $d$-outcome Bell inequality is. Second, the inclusion $MBL_N \subseteq MBL_{N+2}$ seems to hold for $N$ both even and odd in all slices that we studied.

\section*{Acknowledgments.}

We thank Miguel Navascu\'es for discussions, in particular for suggesting the use of Fourier transforms in this context.

This research is supported by by the John Templeton Foundation Grant 60607 ``Many-box locality as a physical principle''; by the Singapore Ministry of Education Academic Research Fund Tier 3 (Grant No. MOE2012-T3-1-009); and by the National Research Fund and the Ministry of Education, Singapore, under the Research Centres of Excellence programme. YZ is also supported by NSFC (Grant Nos. 61671082, 61572081), and the China Scholarship Council. JDB acknowledges support from the Swiss National Science Foundation (SNSF), through the NCCR QSIT and the Grant number PP00P2-150579.

\appendix
\section{Evaluation of $g(\mu)$}\label{appa}

In this section, we provide a detailed description of the evaluation of $g(\mu)$.
 
First, from Eq.~\eqref{eqn:fourierHalf}, it is easy to conclude that
$\mathcal{F}^{-1}_d\left[ \mathcal{J}\right] (k,l)=0$ for $k$ or $l$ even, and
\begin{eqnarray}
\mathcal{F}^{-1}_d\left[ \mathcal{J}\right] (k,l)
&=&\frac{16 \omega^{k+l}}{d^2(1-\omega^{k})(1-\omega^{l})}
\end{eqnarray}
for $k, l$ odd.

Recall the definition of $g(\mu)$ from Eq.~\eqref{eqn:mu}, we have
\begin{eqnarray}
    g(\mu)&=&\frac{16}{d^2 4^{d-1}}\sum_{k,l=odd}\Big((1+\omega^k)(1+\omega^l)+\nonumber\\
    & &(1-\omega^k)(1-\omega^l)\mu\Big)^{d-1} \frac{\omega^{k+l}}{(1-\omega^{k})(1-\omega^{l})}\nonumber\\
    &=&\frac{16}{d^2 4^{d-1}}\sum_{k,l=odd}\sum_{m=0}^{d-1} {{d-1}\choose{m}}\mu^m(1-\omega^k)^{m-1} \nonumber\\ 
    & &(1-\omega^l)^{m-1} (1+\omega^k)^{d-m-1}(1+\omega^l)^{d-m-1} \omega^{k+l}\nonumber.
\end{eqnarray}

Here, we exchange the order of sums, and $k, l$ are symmetric, then we have
\begin{equation}
    g(\mu)=\frac{16}{d^2 4^{d-1}}\sum_{m=0}^{d-1} {{d-1}\choose{m}}\mu^m S^2(d,m),
\end{equation}
where
\begin{equation}
   S(d,m)= \sum_{k=odd}^{d-1} (1-\omega^k)^{m-1} (1+\omega^k)^{d-m-1}\omega^k.
\end{equation}
 
In the following, we aim to simplify $S(d,m)$. First we focus on the case of $m=0$ and expand $(1+\omega^k)^{d-1}$ in the binomial form. Then, we have
 \begin{eqnarray}
     S(d,0)&=&\sum_{k=odd}^{d-1}\sum_{t=0}^{d-1}{{d-1}\choose{t}}\frac{\omega^{k(t+1)}}{1-\omega^k}\nonumber\\
     &=&\sum_{t=0}^{d-1}{{d-1}\choose{t}}\sum_{k=odd}^{d-1}\frac{\omega^{k(t+1)}}{1-\omega^k}.\nonumber
 \end{eqnarray}
 We apply ${{d-1}\choose{t}}={{d-1}\choose{d-t-1}}$, and obtain that
 \begin{eqnarray}
 \label{eqn:S}
    S(d,0) &=&\sum_{t=0}^{d/2-1}{{d-1}\choose{t}}\bigg(\sum_{k=odd}^{d-1}\frac{\omega^{k(t+1)}+\omega^{k(d-t)}}{1-\omega^k}\bigg)\nonumber\\
     &=&\sum_{t=0}^{d/2-1}{{d-1}\choose{t}}\cdot 0\\
     &=&0.\nonumber 
 \end{eqnarray}
The above equation holds since on the one hand the sum of $(\omega^{k(t+1)}+\omega^{k(d-t)})/(1-\omega^k)$ and its conjugate is $0$, which implies that its real part is $0$; on the other hand, its imaginary part is $0$ because $S(d,0)$ must be a real number.
 
Then for $m\geq 1$, we rewrite $(1-\omega^k)^{m-1}$ and $(1-\omega^k)^{d-1-m}$ in the binomial form, then exchange the order of sum. Afterwards, we know that
 \begin{eqnarray}
 \label{eq:s(d,m)}
     S(d,m)&=&\sum_{s=0}^{m-1}\sum_{t=0}^{m-1}(-1)^s{{m-1}\choose{s}}{{d-m-1}\choose{t}}\nonumber\\ 
     & &\bigg(\sum_{k=odd}^{d-1}\omega^{k(s+t+1)}\bigg)\nonumber\\ 
     &=&\sum_{s=0}^{m-1}\sum_{t=0}^{m-1}(-1)^s{{m-1}\choose{t}}{{d-m-1}\choose{s}}\nonumber\\ 
     & &\bigg(\frac{d}{2}\big(\delta_{s+t+1=0}-\delta_{t=d/2-s-1}\big)\bigg)\\
     &=&\frac{d}{2}\sum_{s=0}^{m-1}(-1)^{s+1}{{m-1}\choose{s}}{{d-m-1}\choose{d/2-s-1}},\nonumber
 \end{eqnarray}
Here, Eq. ~\eqref{eq:s(d,m)} holds since 
\begin{equation}
   \sum_{k=odd}^{d-1}\omega^{kt}=\frac{d}{2}\big(\delta_{t=0}-\delta_{t=d/2}\big), 1\leq t\leq d.\nonumber
\end{equation}
 
Now the summation  over $s$ can be evaluated in Mathematica:
 \begin{equation}
    S(d,m)=-\frac{d}{2}\frac{2^{d-2}\Gamma(\frac{d-m}{2})}{\Gamma(\frac{d}{2})\Gamma(1-\frac{m}{2})}.
 \end{equation}
 
Finally, we obtain that
 \begin{equation}
    g(\mu)=\sum_{m=odd}^{N}C(d,m)\mu^m.
 \end{equation}
 
 Third, we introduce two properties of Gamma function to calculate $C(\infty,m)$. One is called as the duplication formula:
 \begin{equation}
 {\Gamma(\eta)}\Gamma(\eta+\frac{1}{2})=2^{1-2\eta}\sqrt{\pi}\Gamma(2\eta). 
 \end{equation}
 Another one is a equivalent condition:
 \begin{eqnarray}
   \lim_{n\rightarrow \infty}\frac{\Gamma(n+\eta)}{\Gamma(n)n^\eta}=1.  
 \end{eqnarray}
 
 Subsequently, we conclude Eq.~\eqref{eq:doubleFactorial} by applying the duplication formula and the equivalent condition. That is,
 \begin{eqnarray}
     & &C(\infty,m)=\lim_{d\rightarrow\infty}C(d,m)\nonumber\\
     &=&\frac{16\Gamma^2(m)}{\Gamma(m+1)\Gamma^2(\frac{m+1}{2})}\lim_{d\rightarrow \infty}\frac{\Gamma(d)\Gamma(d-m)}{4^d \Gamma^2(\frac{d}{2})\Gamma^2(\frac{d-m+1}{2})}\nonumber\\
     &=&\frac{16\Gamma(\frac{m}{2})}{m\sqrt{\pi}2^{1-m}\Gamma(\frac{m+1}{2})}\lim_{d\rightarrow \infty}\frac{\Gamma(\frac{d+1}{2})\Gamma(\frac{d-m}{2})}{ 2^{2+m}\pi\Gamma(\frac{d}{2})\Gamma(\frac{d-m+1}{2})}\nonumber\\
     &=&\frac{2\Gamma(\frac{m}{2})}{m \pi^{3/2} \Gamma(\frac{m+1}{2})}.
 \end{eqnarray}
 
 Note that only odd $m$ needs to be considered. In this way, Eq.~\eqref{eq:doubleFactorial} is obtained by the definition of the Gamma function.
 
%
%
%
%

\end{document}